# Energetic Particle Observations from Parker Solar Probe using Combined Energy Spectra from the ISʘIS Instrument Suite

C.J. Joyce[1], D.J. McComas[1], E.R. Christian[2], N.A. Schwadron[1,4], M.E. Wiedenbeck[4], R.L. McNutt Jr.[5], C.M.S. Cohen[6], R.A. Leske[6], R.A. Mewaldt[6], E.C. Stone[6], A.W. Labrador[6], A.J. Davis[6], A.C. Cummings[6], D.G. Mitchell[5], M. E. Hill[5], E.C. Roelof[5], J.R. Szalay[1], J. S. Rankin[1], M.I. Desai[7,8], J. Giacalone[9], W.H. Matthaeus[10].

[1]*Department of Astrophysical Sciences, Princeton University, Princeton, NJ 08540, USA*

[2]*Goddard Space Flight Center, Greenbelt, MD 20771, USA*

[3]*University of New Hampshire, Durham, NH 03824, USA*

[4]*Jet Propulsion Laboratory, California Institute of Technology, Pasadena, CA 91109, USA*

[5]*Johns Hopkins University Applied Physics Laboratory, Laurel, MD 20723, USA*

[6]*California Institute of Technology, Pasadena, CA 91125, USA*

[7]*Southwest Research Institute, San Antonio, TX 78228, USA*

[8]*University of Texas at San Antonio, San Antonio, TX 78249, USA*

[9]*University of Arizona, Tucson, AZ 85721, USA*

[10]*University of Delaware, Newark, DE 19716, USA*

ABSTRACT

The Integrated Science Investigations of the Sun (ISʘIS) instrument suite includes two Energetic Particle instruments: EPI-Hi, designed to measure ions from ~1-200 MeV/nuc, and EPI-Lo, designed to measure ions from ~20 keV/nuc to ~15 MeV/nuc. We present an analysis of eight energetic proton events observed across the energy range of both instruments during PSP's first two orbits in order to examine their combined energy spectra. Background corrections are applied to help resolve spectral breaks between the two instruments and are shown to be effective. In doing so we demonstrate that, even in the early stages of calibration, ISʘIS is capable of producing reliable spectral observations across broad energy ranges. In addition to making groundbreaking measurements very near the Sun, IS☉IS also characterizes energetic particle populations over a range of heliocentric distances inside 1 au. During the first two orbits, IS☉IS observed energetic particle events from a single corotating interaction region (CIR) at three different distances from the Sun. The events are separated by two Carrington rotations and just 0.11 au in distance, however the relationship shown between proton intensities and proximity of the spacecraft to the source region shows evidence of the importance of transport effects on observations of energetic particles from CIRs. Future IS☉IS observations of similar events over larger distances will help disentangle the effects of CIR-related acceleration and transport. We apply similar spectral analyses to the remaining five events, including four that are likely related to stream interaction regions (SIRs) and one solar energetic particle (SEP) event.

## 1. INTRODUCTION

As part of NASA's Parker Solar Probe (PSP) mission (Fox et al. 2016), the Integrated Science Investigation of the Sun (IS☉IS) (McComas et al. 2016) instrument suite was designed to provide a comprehensive characterization of the near-Sun energetic particle environment. The IS☉IS instrument suite includes two Energetic Particle Instruments: EPI-Hi, designed to measure ions from ~1-200 MeV/nuc, and EPI-Lo, designed to measure ions from ~20 keV/nuc to ~15 MeV/nuc. In just the first two PSP orbits, IS☉IS has already provided fascinating observations of the energetic particle environment near the Sun (McComas et al. 2019). As more data is collected, IS☉IS will provide new insights into the sources, acceleration and transport of energetic particles in the inner heliosphere. Characterization of acceleration mechanisms will be greatly facilitated by the ability to observe SEP events much closer to their acceleration sites near the Sun, while comparisons with spacecraft at 1 au will deepen our understanding of the role of transport in shaping SEP events.

Another important aspect of IS☉IS observations is that it will be able to measure energetic particle events not just close to the Sun, but over a variety of distances inside 1 au due to PSP's highly elliptical orbits. This will be critical for understanding the role of transport effects on CIR observations; as IS☉IS will be able to measure individual CIRs at a range of distances from the source region across multiple Carrington rotations.

Suprathermal seed populations are expected to play a major role in the development of large SEP events (Mason et al. 1999a; Desai et al. 2006; Mewaldt et a. 2012a). IS⊙IS will be able to test these assertions for the first time by measuring ambient suprathermal intensities near the Sun and also by measuring small SEP events near the Sun, which may occur frequently but dissipate before they can be observed at 1 au. Such never before measured events may play a critical role in seeding the solar wind.

During PSP's first two orbits, IS⊙IS measured a variety of different energetic particle events, some observed only by EPI-Hi, some only by EPI-Lo and some by both instruments (McComas et al. 2019). EPI-Hi and EPI-Lo were specifically designed with overlapping energy coverage, enabling cross-calibration between the two instruments and determination of the energy spectrum across the full energy range of the IS⊙IS suite (McComas et al. 2016). This study represents the first detailed cross-instrument spectral analysis of IS⊙IS data from multiple energetic particle events. Here, we analyze eight energetic proton events observed by both instruments in order to evaluate the consistency of their observations and to analyze the events using their combined energy spectra. These events include three related to a single CIR, observed from three different heliospheric distances, four events that are likely SIR-related and one SEP event observed very near perihelion. Section 2 describes the methods used to reduce the impact of background (i.e. signal not associated with the events studied) on the event spectra, facilitating more accurate event analysis. Section 3 provides an overview of the spectral characteristics of the eight events and evaluates the consistency of the spectra of the two instruments. Section 4 discusses the physical properties of these events in the context of previous observations by spacecraft at 1 au. Section 6 presents conclusions. The Appendix includes an overview of the instruments and discusses sources of background for EPI-Hi and EPI-Lo.

## 2. BACKGROUND SUBTRACTION

Energetic particle events are typically characterized by spectra that are smooth and continuous across energy ranges and spectral slopes which are harder (flatter) at low energies and softer (steeper) at high energies. The energy range at which the spectrum transitions from hard to soft is referred to as a spectral break or rollover. The energies at which these breaks occur is determined by the acceleration mechanism involved (Desai et al. 2016), typically shocks for CMEs (Desai & Giacalone 2016) or corotating interaction regions (Gosling 1996), and reconnection for impulsive events usually associated with solar flares (Mason 2007). Spectral indices below the break are generally also related to the acceleration mechanism. Spectral break energies, as well as the spectral index above the break, have also been shown to be related to the rigidity dependence of particle scattering mean free paths (Li, Hu & Zank 2005; Li et al. 2009; Schwadron et al. 2015).

When spectral breaks occur near the overlapping energy region of two instruments, it is important to ensure that the difference in spectral indices is physical and not due to instrumental effects. For most of the events studied here, the differences in spectral slopes between EPI-Hi and EPI-Lo indicate the presence of a spectral break just below the lower energy limit of EPI-Hi, which is difficult to resolve due to the effects of instrumental background on the EPI-Lo spectrum. We therefore apply a correction method to reduce the amount of background in the spectra. In this paper we define background as any sources that have a persistent effect on the instrument measurements, both during and outside events. This includes instrumental background due to false signals, such as cross-talk and accidentals caused by ultraviolet (UV) photons and galactic cosmic rays (GCRs) for EPI-Lo, and background due to real signals, such as cosmic rays (CRs) and solar/heliospheric (SH) particles for EPI-Hi (see the Appendix for more details). Since this paper is concerned with event analysis and comparing the EPI-Hi and EPI-Lo spectra produced by a common signal associated with those events, it is desirable to reduce all of these backgrounds as much as possible for this study.

Since the sources of background with which we are concerned are present at all times and vary slowly in time, we quantify them by examining quiet periods, defined here as intervals when both EPI-Hi and EPI-Lo observe steady conditions with no evidence of an event in progress. To maximize counting statistics while also accounting for potential temporal variations of the background we apply the following methods: (1) We take all quiet-time periods in the data set and use them to produce time-averaged spectra for EPI-Hi and EPI-Lo that are characteristic of the background typically observed. (2) We scale the averaged background spectra for each event based on the average particle intensities during a specified quiet interval preceding the event. (3) We subtract the scaled quiet-time spectrum from the events spectrum.

Figure 1 shows the time-averaged flux spectra from 13 quiet periods during the first two PSP orbits. While there is some variation from interval to interval, the features of the average spectra shown in Figure 1 are generally characteristic of the quiet-time spectra throughout the dataset, with differences primarily being in overall magnitude, along with some statistical variation. Figure 2 shows an example of the effects of quiet-time subtraction for an event that occurred from day of year (DOY) 44 to 49 of 2019 (referred to here as Event 4 or E4). The average count rate spectrum used to generate Figure 1 was scaled based on the average proton count rates summed over all energies during the preceding quiet interval (from 8:00 on DOY 41 to 18:00 on DOY 44) and subtracted from the average count rate spectra during Event 4 before being converted to flux.

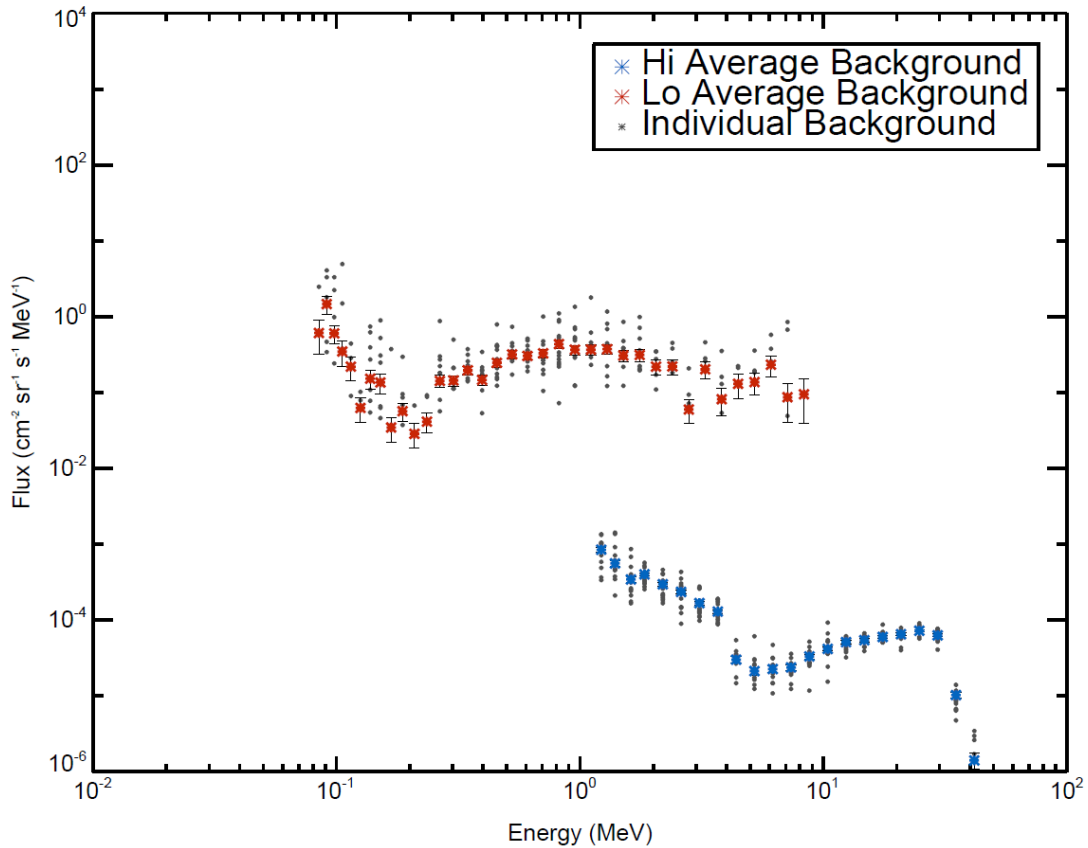

Figure 1: Average quiet-time differential flux spectra for EPI-Hi (blue) and EPI-Lo (red) over the course of the first two orbits of PSP. The quiet intervals used are specified in Table A1 in the Appendix. The error bars shown are statistical uncertainties. The small gray points in the background are the individual spectra for all of the quiet intervals used, showing that the average spectra are representative of the general background characteristics. In the EPI-Hi spectrum, we can see a transition at about 6 MeV between the lower energy range dominated by SH particles and the higher energy dominated by CRs (see Appendix for details on background sources).

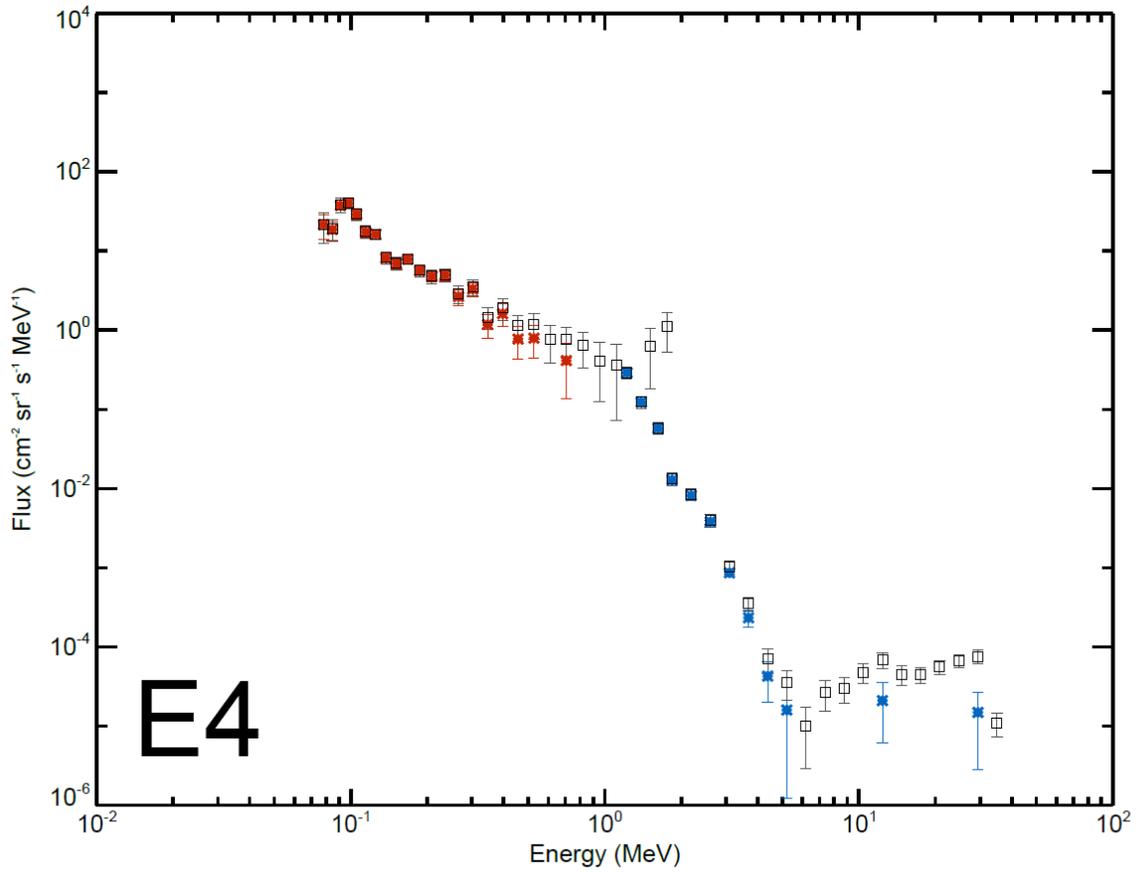

Figure 2: Comparison of unaltered spectrum (open black squares) to the quiet-time corrected spectrum for Event 4 (asterisks). The modified spectra are shown in blue for EPI-Hi and red for EPI-Lo. Points that become negative or have errors larger than the value of the point following background subtraction have been excluded from the plot. The quiet time intervals used to scale the background subtraction for each event are specified by Table A2 in the Appendix, along with start/end times for each event.

Figure 2 shows how the background sources can impact the comparison between EPI-Hi and EPI-Lo during an event. The unmodified spectrum shows a relatively constant spectral slope above ~0.1, however with the background subtracted, we see that the spectrum actually begins to soften at ~0.3 MeV. This is especially important for comparing the spectra of the two instruments as the background sources which are shown to dominate the EPI-Lo spectrum near 1 MeV are not at all related to the energetic particles measured by EPI-Hi near 1 MeV. The background correction to EPI-Hi has less bearing on the comparison but does result in a slightly softer spectrum. Some spectral features, such as the steeper spectral slope seen for EPI-Lo between ~0.09-0.15 MeV and the dip at ~2 MeV in the EPI-Hi spectrum, are likely not physical and will be corrected as ongoing calibration efforts improve. Additionally, the increased difficulty in estimating detector efficiencies at lower energies results in a non-physical turndown in the EPI-Lo spectrum at low energies. This analysis takes place early in the mission while instrument calibrations are still ongoing. As more data is collected and analyzed, these observations will be improved through refinement of instrument parameters and data analysis.

## 3. SPECTRAL COMPARISON

During the first two PSP orbits, IS☉IS observed a wide variety of energetic particle events. Some of these events were observed only by EPI-Lo, such as the CME-associated event on DOY 315 of 2018 (McComas et al. 2019; Giacalone et al. 2019; Mitchell et al. 2019), which presumably lacked acceleration mechanisms strong enough to accelerate significant numbers of particles up to the energies measurable by EPI-Hi. Other events were observed only by EPI-Hi, such as the two CIR-related events analyzed by McComas et al. 2019 and four of the seven stream interaction region (SIR) events studied by Cohen et al. (2019) — likely due to transport effects preventing low energy particles from reaching the spacecraft. Here we analyze only events that showed particle intensities in both EPI-Hi and EPI-Lo that were significantly above background rates. We have also rejected any events in which one or both of the instruments was powered off in the middle of the event (such as the 21 April, 2019 event studied by Schwadron et al. (2019) and Wiedenbeck et al. (2019)).

Figure 3 shows the spectra for all eight events analyzed here and Table 1 shows their event-averaged fluxes and spectral characteristics. All of the events show significant differences in the spectral indices at high and low energies. Six of the eight events (E1, E3, E4, E6, E7, E8) are very soft at high energies with spectral indices less than -5 and much harder at lower energies. These are all consistent with SIRs or CIRs (see further discussion in Section 4). Two of the events (E2 and E5) are significantly harder at higher energies and somewhat softer than average at lower energies, resulting in somewhat less significant spectral breaks between the two instruments. E3 is also relatively soft at low energies, but at high energies it is similar to the rest of the SIR/CIR-related events.

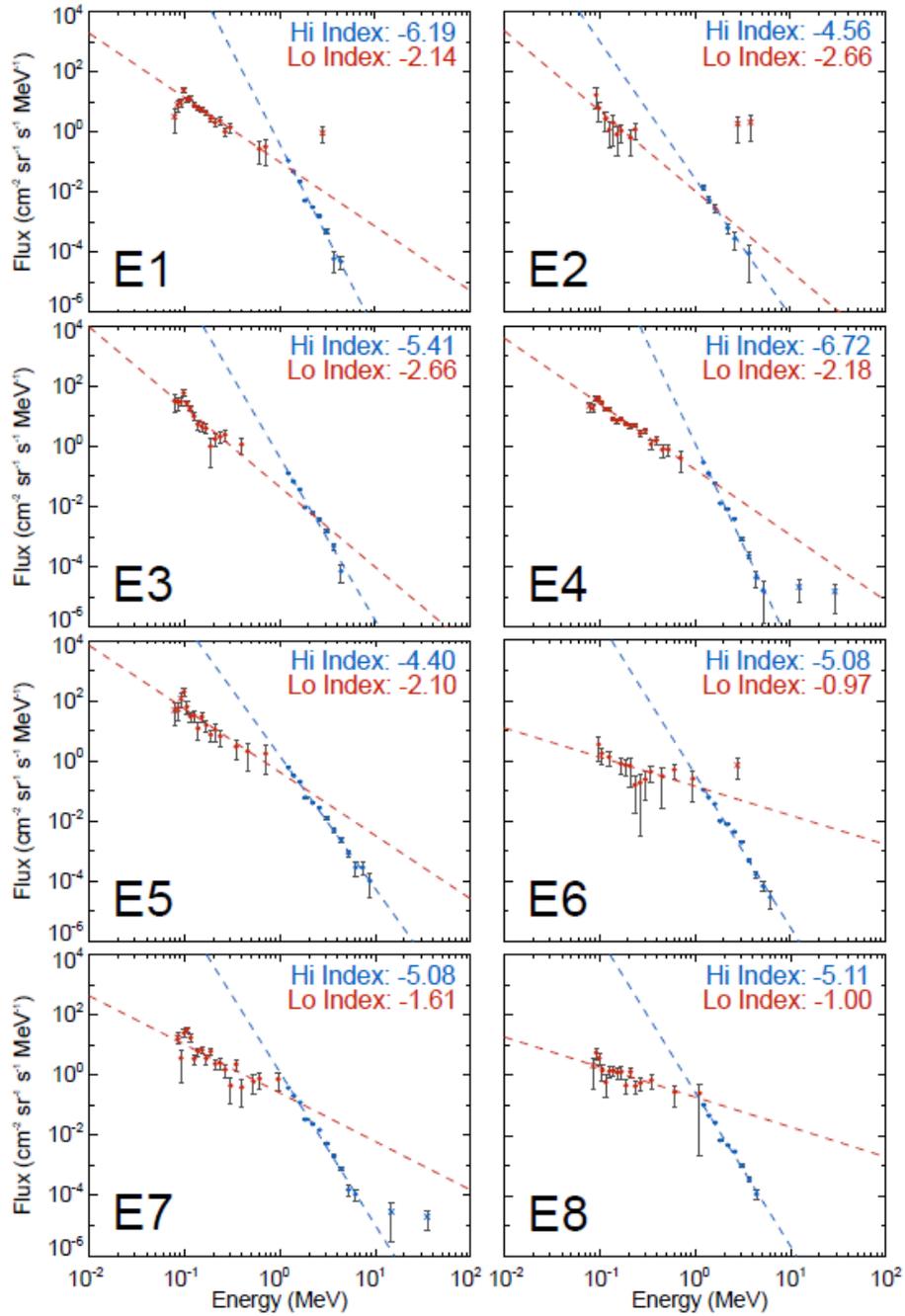

Figure 3: Combined differential flux spectra for all events shown here, with background correction applied. Power law fits are applied to the EPI-Hi (blue) and EPI-Lo (red) spectra to aid in the comparison. For each event spectra, only the data points represented by diamond symbols are used for the power law fit, which excludes spectral turnups at higher energies due to remaining cross-talk background for EPI-Lo and cosmic rays for EPI-Hi, as well as the turndown at lower energies (points excluded by the fits are represented with X's).

<div align="center">

**Table 1**

Event Data

</div>

| Event # | Start YYYY-DOY | R (au) | Average Hi Flux (cm$^{-2}$ sr$^{-1}$ s$^{-1}$ MeV$^{-1}$) | Average Lo Flux (cm$^{-2}$ sr$^{-1}$ s$^{-1}$ MeV$^{-1}$) | Hi Index | Lo Index | Index Diff. |
|---|---|---|---|---|---|---|---|
| 1 | 2018-320 | 0.38 | $5.84 \times 10^{-3} \pm 1.99 \times 10^{-4}$ | $1.55 \times 10^{0} \pm 1.08 \times 10^{-1}$ | $-6.19 \pm 0.80$ | $-2.14 \pm 0.46$ | 4.05 |
| 2 | 2019-021 | 0.94 | $1.63 \times 10^{-3} \pm 3.19 \times 10^{-4}$ | $8.99 \times 10^{-1} \pm 2.43 \times 10^{-1}$ | $-4.56 \pm 1.07$ | $-2.66 \pm 1.10$ | 1.9 |
| 3 | 2019-031 | 0.92 | $9.82 \times 10^{-3} \pm 4.42 \times 10^{-4}$ | $3.81 \times 10^{0} \pm 3.76 \times 10^{-1}$ | $-5.41 \pm 0.80$ | $-2.66 \pm 0.57$ | 2.75 |
| 4 | 2019-045 | 0.85 | $1.40 \times 10^{-2} \pm 3.03 \times 10^{-4}$ | $3.29 \times 10^{0} \pm 1.63 \times 10^{-1}$ | $-6.72 \pm 0.67$ | $-2.18 \pm 0.39$ | 4.54 |
| 5 | 2019-094 | 0.17 | $2.92 \times 10^{-2} \pm 1.24 \times 10^{-3}$ | $1.07 \times 10^{1} \pm 1.45 \times 10^{0}$ | $-4.40 \pm 0.44$ | $-2.10 \pm 0.48$ | 2.3 |
| 6 | 2019-144 | 0.87 | $7.68 \times 10^{-3} \pm 2.43 \times 10^{-4}$ | $4.17 \times 10^{-1} \pm 1.04 \times 10^{-1}$ | $-5.08 \pm 0.58$ | $-0.97 \pm 0.43$ | 4.11 |
| 7 | 2019-171 | 0.94 | $2.63 \times 10^{-2} \pm 7.49 \times 10^{-4}$ | $2.09 \times 10^{0} \pm 1.93 \times 10^{-1}$ | $-5.08 \pm 0.58$ | $-1.61 \pm 0.36$ | 3.47 |
| 8 | 2019-197 | 0.83 | $5.48 \times 10^{-3} \pm 2.06 \times 10^{-4}$ | $4.97 \times 10^{-1} \pm 7.32 \times 10^{-2}$ | $-5.11 \pm 0.80$ | $-1.00 \pm 0.38$ | 4.11 |

**Notes.** Exact start/end times used for each event are specified in Table A2. EPI-Hi and EPI-Lo fluxes are time-averaged and integrated over 1-10 MeV and 60-800 keV, respectively, following background subtraction.

## 4. DISCUSSION

The importance of background subtraction in this study is motivated by how small the energetic particle events observed thus far at PSP have been. Spectral analyses of energetic particle events tend to focus on large gradual events and heavy ions rather than small, relatively soft proton events like those studied here. These larger events, typically CME-driven, are capable of accelerating large numbers of particles to high energies and are characterized by hard spectral indices and high spectral break energies often >10 MeV (Mewaldt et al. 2012b). These events are characterized by smooth, continuous spectra across the overlapping energy range at intensities well above background levels. Future IS⊙IS observations of such large events will present ideal opportunities to refine the cross-calibration between EPI-Hi and EPI-Lo.

Three of the events shown here (E1, E3 and E4) have been identified as SIRs by Cohen et al. (2019), though E1 followed the passage of a CME two days prior and it is possible this event originates from a shock formed by the CME after it passed PSP (McComas et al. 2019). SIR-related events are observed when the spacecraft passes through a flux tube connected to the source region that has been filled with energetic particles. Thus, these events are characterized by a distinctive lack of anisotropy or dispersion, features commonly associated with remote-source SEPs.

When SIRs persist for more than one solar rotation, they are referred to as CIRs and may be measured multiple times by a single spacecraft as it passes through magnetic field lines connected to the source region. E6, E7 and E8 all appear to be associated with a single CIR, with each event being separated in time by approximately one Carrington rotation. Figure 4 shows the alignment of the solar footpoint longitudes over this time period, which further solidifies this interpretation. Across the three events, the spectral index for EPI-Hi remains constant at about -5.1, while the EPI-Lo spectral index becomes softer at greater distances. Interestingly, despite the softening of the spectra at low energies, the EPI-Hi average flux increases with distance from the Sun. Assuming the heliocentric distance of the spacecraft is indicative of the proximity of the spacecraft to the source region, the softening of the spectrum at low energies is in keeping with the CIR shock acceleration theory of Fisk & Lee 1980 (Mason et al. 1999b), though the fluxes at high energies are the opposite of what would be expected. Nonetheless, the three events are only separated by 0.11 au and it is possible that the time evolution of the interaction region or the spacecraft's connection to it also play a significant role in the differences between events.

E2 is somewhat unique compared with the other events. The event is clearly not SEP-related as it lacks significant anisotropy or dispersion, however its somewhat more linear spectrum also sets it apart from the other SIR-related events shown here. E2 is preceded by four days of EPI-Hi activity during which no significant activity above background is measured by EPI-Lo. As mentioned previously, IS☉IS has observed several SIR events that were observed by EPI-Hi and not EPI-Lo. Based on these characteristics, it would seem possible that PSP was connected to an SIR for the first four days (pre-E2) and on the fourth day became connected to a different source coinciding with the rise in intensities measured by EPI-Lo. Figure 5 shows the close alignment of the solar footpoint longitudes between E2 and E4, which suggests that E2 and E4 may be related to a common CIR, however given the significant differences in spectral shapes and overall intensities this seems unlikely.

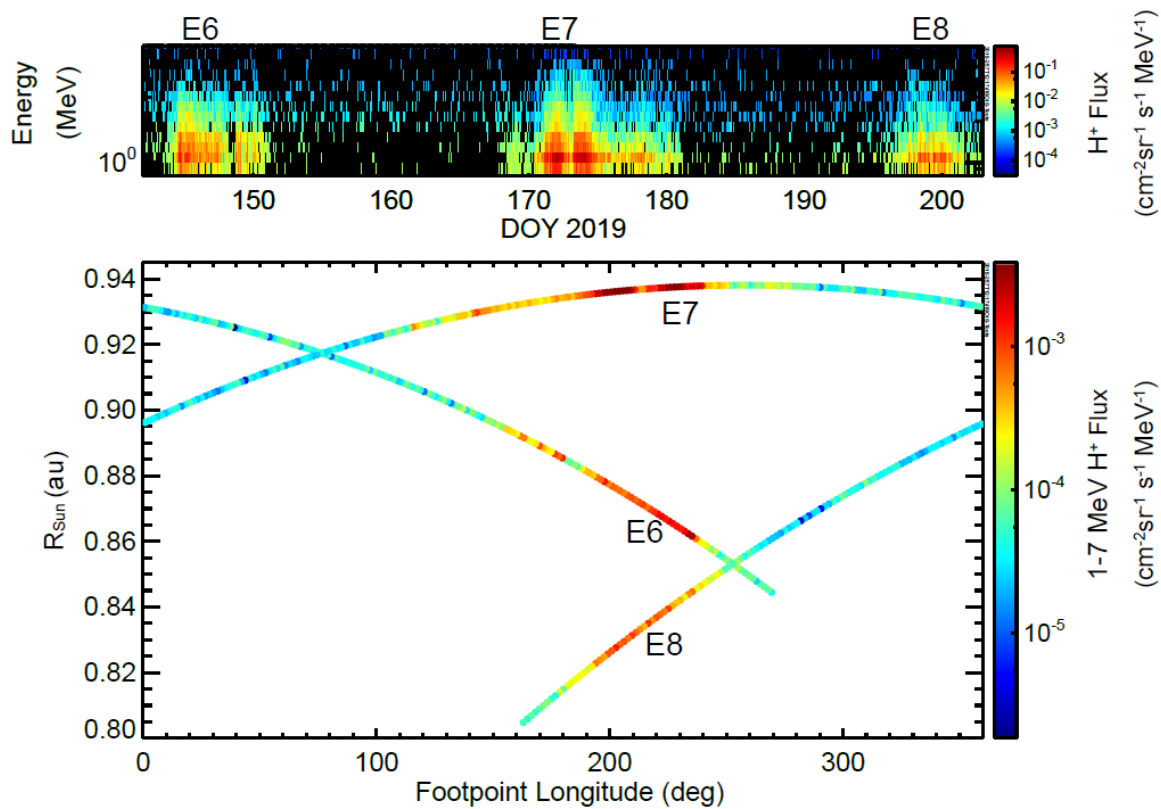

Figure 4: EPI-Hi LET1 spectrogram (top panel) and EPI-Hi LET1 flux integrated over 1-7 MeV as a function of radial distance from the Sun and magnetic footpoint on the Sun (assuming nominal Parker spiral with solar wind speed of 400 km/s). Events E6, E7 and E8 are labelled in both panels.

We use global heliospheric modelling to examine what might have produced these unusual observations. Figure 6 shows plasma densities simulated by the ENLIL model during E2, E4 and the pre-E2 active period observed only by EPI-Hi. The pre-E2 period appears to be associated with a fairly weak compression region passing over the spacecraft. This compression is then followed from behind by a second somewhat weaker compression region. As it passes over PSP, this second compression region appears to overtake the first, forming a stronger compression. The formation of this merged compression seems to coincide with the rise in EPI-Lo intensities, which seems to indicate that it is the source of energetic particles for E2. The merged compression is followed by a rarefaction region, which likely stretches the magnetic field in a more radial direction, allowing PSP to remain connected to the acceleration source even after it has passed over the spacecraft. This explains why particle intensities remain high for about one day after the merged compression has passed over PSP. One solar rotation later, the simulation indicates that the merged compression has strengthened significantly, possibly forming a shock beyond 1 au. As a result, E4 is characterized by much higher fluxes and a much more typical SIR spectrum.

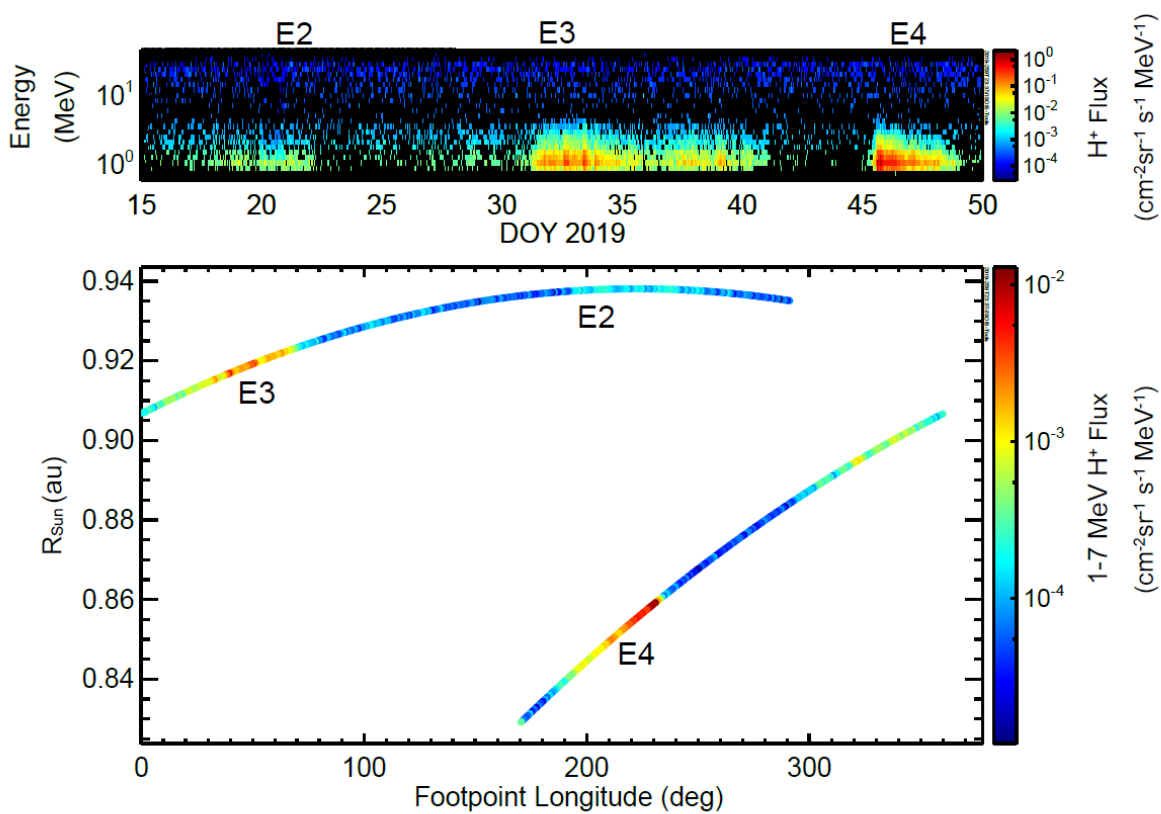

Figure 5: EPI-Hi LET1 spectrogram (top panel) and EPI-Hi LET1 flux integrated over 1-7 MeV as a function of radial distance from the Sun and magnetic footpoint on the Sun (assuming nominal Parker spiral with solar wind speed of 400 km/s). Events E2, E3 and E4 are labelled in both panels.

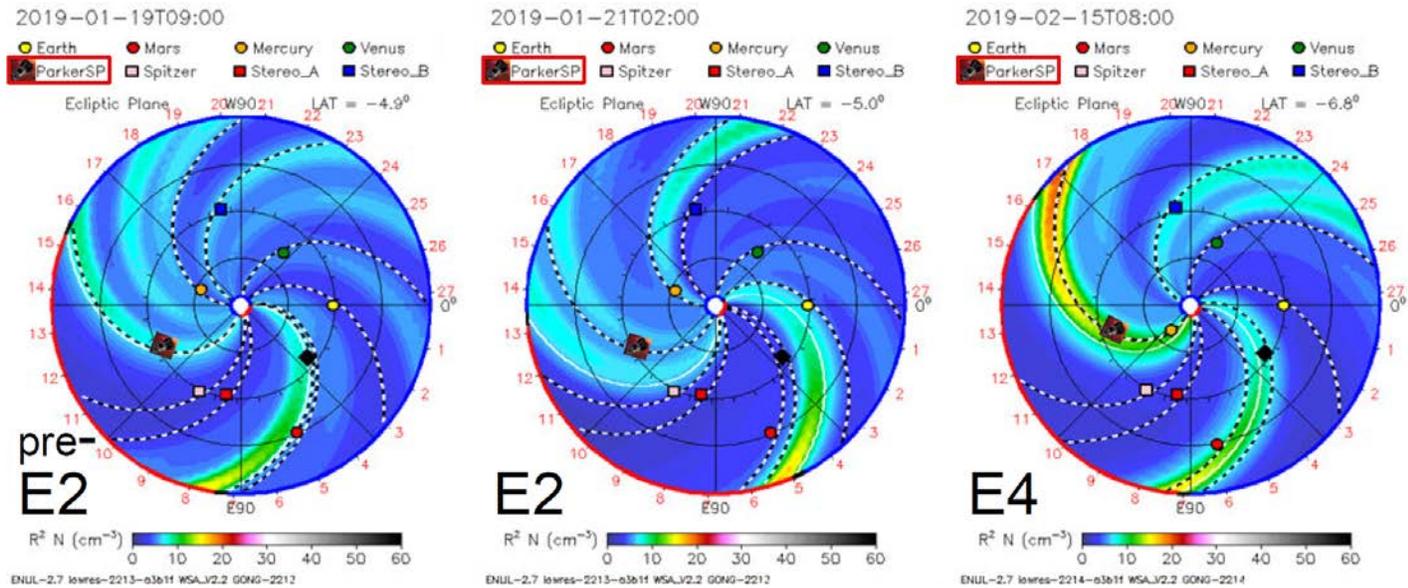

Figure 6: Plasma densities in the inner heliosphere simulated by the ENLIL model for E2, E4 and about two days prior to E2 when only EPI-Hi was observing significant activity.

Despite seemingly being related to an SIR, the more linear spectrum of E2 sets it apart from the other SIR/CIR-related events. This may be evidence for local acceleration occurring within the merged compression region associated with the event. Mason et al. (2000) analyzed ACE observations of energetic particles near CIRs and interpreted the lack of a low-energy rollover as evidence of local acceleration. And, Giacalone et al. (2002) presented a model of particle acceleration at a gradual plasma compression across a CIR in which particles are accelerated locally at the compression, resulting in more linear spectra. It is possible that the closer spectral indices at high and low energies for E2 may be indicative of local acceleration playing a more substantial role in producing the energetic particle population early in the formation of the merged compression region. By the time the SIR is observed again at E4, the spectrum is presumably dominated by particles accelerated remotely where the compression has strengthened beyond 1 au.

Seven of the eight events shown here have been shown to be SIR/CIR-related, however E5 is clearly an impulsive SEP event (for details, see McComas et al. 2019 or Leske et al. 2019) that occurred near aphelion. This event is quite interesting in that such a small event observed just 0.17 au from the Sun would probably be impossible to measure at 1 au. Such small impulsive events may occur frequently in the inner heliosphere and play an important role in providing seed populations to be accelerated by subsequent larger, gradual SEP events.

In summary, six of the eight events shown here are clearly SIR/CIR related, with spectra that are fairly typical of such events. Richardson et al. (2004) found that CIR-related events observed at 1 au tended to have spectral indices of about -2 below ~1 MeV and about -4 above. The six CIR/SIR-associated events shown here are similar in overall shape, but are softer on average for EPI-Hi (-5.60 ± 0.63) and harder on average for EPI-Lo (-1.76 ± 0.63). This difference with observations at 1 au is again in keeping with the transport theory of Fisk & Lee 1980.

## 5. CONCLUSIONS

One year after PSP first began its journey to the Sun, early results show IS☉IS making a number of exciting observations in a never before explored region of space. We have used these new data to provide the first detailed analysis of multiple energetic particle events observed by IS☉IS using the combined EPI-Hi/EPI-Lo spectrum. All of the events observed by IS☉IS thus far have been relatively weak and it is perhaps not surprising that most of these events show spectral breaks between EPI-Hi and EPI-Lo at relatively low energies (~0.3-1.0 MeV). For the events shown here, the unmodified EPI-Lo spectrum in the intermediate energy range where the spectral breaks are seen is largely dominated by background. We have shown that our simple background subtraction method is somewhat effective in resolving spectral breaks in the EPI-Lo spectrum in the intermediate energy range, however a more detailed analysis of quiet-time spectra along with further improved characterization of instrumental efficiencies/response should lead to even better results in the future. We have demonstrated how IS☉IS observations will present a unique opportunity to study the role of CIR/SIR particles in the innermost heliosphere and to provide observations of individual CIRs at a range of distances over multiple Carrington rotations due to the highly eccentric orbit of PSP. This, combined with the opportunities for multi-spacecraft observations with both Solar Orbiter and 1 au spacecraft, should help to disentangle the effects of acceleration and transport on the spectra measured by IS☉IS during CIR-related events. We look forward to many years of IS☉IS providing a comprehensive characterization of the near-Sun energetic particle environment over a broad range of particle energies.

ACKNOWLEDGEMENTS

This work was supported as a part of the Integrated Science Investigations of the Sun on NASA's Parker Solar Probe mission, under contract NNN06AA01C. The ISʘIS data and visualization tools are available to the community at: https://spacephysics.princeton.edu/missions-instruments/isois; data are also available via the NASA Space Physics Data Facility (https://spdf.gsfc.nasa.gov/).Simulation results have been provided by the Community Coordinated Modeling Center at Goddard Space Flight Center through their public Runs on Request system (http://ccmc.gsfc.nasa.gov). The ENLIL Model was developed by D. Odstrcil at the Univ. of Colorado at Boulder.

APPENDIX: OVERVIEW OF INSTRUMENTS AND SOURCES OF BACKGROUND

The EPI-Hi instrument is comprised of three telescopes: a double-ended High Energy Telescope (HET) aligned roughly along the nominal Parker spiral direction at 0.25 au, a double-ended Low Energy Telescope (LET1) also roughly aligned along the nominal Parker spiral direction, but 25° offset from HET, and a second, single-ended Low Energy Telescope (LET2), pointed in the ram direction perpendicular to LET1 (McComas et al. 2016, Wiedenbeck et al. 2017). These alignments assume nominal spacecraft pointing. Each telescope has a 45° half-angle conical field of view centered along the plane of the spacecraft's orbit (as illustrated by Figure 18 of McComas et al. 2016). Building upon the heritage of the HET and LET instruments on the STEREO spacecraft (von Rosenvinge et al. 2008; Mewaldt et al. 2008), each telescope includes an array of solid-state detectors (SSDs) that measure the energy deposited by energetic particles. This enables the determination of particle composition via dE/dx vs. total energy techniques that separate different species based on their adherence to characteristic energy deposition tracks. In this analysis, we use measurements of protons made by sunward-facing side of LET1.

EPI-Lo measures the energy and composition of energetic particles using a series of 80 apertures distributed over a 2π steradian field of view, enabling high-resolution characterization of particle angular distributions (see Figures 15 and 19 of McComas et al 2016 for mechanical design and field of view diagram). EPI-Lo is a time of flight (TOF) instrument that also incorporates SSDs which measure the total energy of deposited particles, enabling the determination of particle mass with enough resolution to separate protons, 3He, 4He and heavy ions (see Hill et al. 2017). In this analysis, we use only triple-coincidence (start/stop time and energy) measurements of protons in order to provide the best comparison to the EPI-Hi proton measurements and the highest energy resolution available. To provide suitable comparison with the EPI-Hi measurements, EPI-Lo fluxes are averaged over the apertures from its three sunward-facing wedges, but excluding the five apertures (25, 31, 34, 35 and 44) which are most affected by UV background.

Spectral comparisons between EPI-Hi and EPI-Lo during energetic particle events are complicated by the influence of background particles that affect the observations. Here we are concerned only with instrument signal related to the events being studied, and thus define background as contributions to measured intensities that are consistently present both during and outside of events, whether caused by instrumental effects (such as accidentals or cross-talk) or other ambient energetic particle sources. Ideally, cross-calibration would be done with an a priori known, single, uniform energetic particle source, however interplanetary space is characterized by a variety of unknown sources and effects, making event analysis challenging. Because of the differences in how the two instruments operate and the energy ranges which they measure, the background sources affecting their spectra are different. Section 2 describes the methods used here to correct for instrument background using measurements taken during quiet periods over the course of the entire dataset (defined in Table A1). We note that the fluxes shown for each interval in Table A1 are not well correlated with radial distance from the Sun for either EPI-Hi or EPI-Lo. Table A2 shows the start/end times for each event and the quiet interval used to scale the background subtraction.

In the case of EPI-Hi, the primary energetic particle sources competing with event signals are CRs and SH particles. CRs, which include both GCRs and anomalous cosmic rays (ACRs), are high energy particles originating outside the heliosphere whose intensities in interplanetary space vary slowly with the solar cycle. While CRs lose energy as they penetrate through the heliosphere, CR intensity variations over distances that are relatively short compared to the scale of the heliosphere are likely small, though this has not yet been confirmed by measurement. SH refer to energetic particles 1-10 MeV of solar origin that are consistently present in the interplanetary medium (Logachev, Kecskeméty & Zeldovich, 2002) and decrease with heliocentric distance (Kecskeméty, Müller-Mellin, and Kunow, 2001).

For EPI-Lo, CRs have an impact on observations in a different way. EPI-Lo is not designed to measure CRs, with its upper energy range not extending far into their domain, however CRs (likely protons >50 MeV) can create cross-talk between the start and stop measurements within the instrument. Another consistent source of instrumental background in EPI-Lo data is ultraviolet (UV) photons from the Sun. A small percentage of UV photons are capable of triggering starts and/or stops, which can interfere with particle measurements thereby creating background over a wide range of energies. Using triple coincidence data greatly reduces the background from UV, however coincidence conditions will still be met some small percentage of the time. To reduce the effect of UV, we do not use measurements from the five apertures most affected, though this background can be present at many look directions since the UV that produces background is scattered off electrons or dust particles. Hill et al. (2019) provides a detailed discussion of EPI-Lo instrumental background.

**Table A1**

Quiet-Time Data

| Start Time: | End Time: | R (AU) | Average Hi Flux (cm$^{-2}$ sr$^{-1}$ s$^{-1}$ MeV$^{-1}$) | Average Lo Flux (cm$^{-2}$ sr$^{-1}$ s$^{-1}$ MeV$^{-1}$) |
|---|---|---|---|---|
| 2018-296T14:00:00 | 2018-314T16:00:00 | 0.23 | $1.44\times10^{-4} \pm 8.67\times10^{-6}$ | $2.90\times10^{-1} \pm 2.96\times10^{-2}$ |
| 2018-325T08:00:00 | 2018-327T00:00:00 | 0.47 | $1.63\times10^{-4} \pm 3.02\times10^{-5}$ | $3.92\times10^{-1} \pm 1.58\times10^{-1}$ |
| 2019-007T00:00:00 | 2019-011T12:00:00 | 0.92 | $1.08\times10^{-4} \pm 1.40\times10^{-5}$ | $4.29\times10^{-1} \pm 7.43\times10^{-2}$ |
| 2019-023T00:00:00 | 2019-026T15:00:00 | 0.94 | $1.45\times10^{-4} \pm 1.89\times10^{-5}$ | $2.65\times10^{-1} \pm 6.08\times10^{-2}$ |
| 2019-041T08:00:00 | 2019-044T18:00:00 | 0.87 | $8.05\times10^{-5} \pm 1.45\times10^{-5}$ | $3.50\times10^{-1} \pm 6.58\times10^{-2}$ |
| 2019-052T02:00:00 | 2019-054T12:00:00 | 0.8 | $8.33\times10^{-5} \pm 1.92\times10^{-5}$ | $2.61\times10^{-1} \pm 6.65\times10^{-2}$ |
| 2019-056T00:00:00 | 2019-057T10:00:00 | 0.77 | $9.15\times10^{-5} \pm 2.58\times10^{-5}$ | $8.23\times10^{-1} \pm 2.25\times10^{-1}$ |
| 2019-068T07:00:00 | 2019-073T00:00:00 | 0.61 | $8.41\times10^{-5} \pm 1.24\times10^{-5}$ | $3.12\times10^{-1} \pm 6.49\times10^{-2}$ |
| 2019-074T12:00:00 | 2019-076T12:00:00 | 0.53 | $1.16\times10^{-4} \pm 2.34\times10^{-5}$ | $4.25\times10^{-1} \pm 1.11\times10^{-1}$ |
| 2019-085T02:00:00 | 2019-092T06:00:00 | 0.26 | $7.97\times10^{-5} \pm 9.43\times10^{-6}$ | $2.73\times10^{-1} \pm 4.45\times10^{-2}$ |
| 2019-139T02:00:00 | 2019-143T06:00:00 | 0.84 | $1.43\times10^{-4} \pm 1.93\times10^{-5}$ | $3.76\times10^{-1} \pm 8.34\times10^{-2}$ |
| 2019-152T06:00:00 | 2019-167T10:00:00 | 0.93 | $8.09\times10^{-5} \pm 6.53\times10^{-6}$ | $1.84\times10^{-1} \pm 1.97\times10^{-2}$ |
| 2019-181T09:00:00 | 2019-191T18:00:00 | 0.9 | $9.82\times10^{-5} \pm 9.75\times10^{-6}$ | $2.52\times10^{-1} \pm 2.97\times10^{-2}$ |

**Notes.** EPI-Hi and EPI-Lo fluxes are time-averaged and integrated over 1-10 MeV and 60-800 keV, respectively.

**Table A2**

Start/Stop Times for Events and Preceding Quiet Periods

| Event # | Start Time: | End Time: | Quiet-time Start: | Quiet-time End: |
|---|---|---|---|---|
| 1 | 2018-320T00:00:00 | 2018-323T16:00:00 | 2018-311T00:00:00 | 2018-314T16:00:00 |
| 2 | 2019-021T05:00:00 | 2019-022T05:00:00 | 2019-007T00:00:00 | 2019-011T12:00:00 |
| 3 | 2019-031T05:00:00 | 2019-032T15:00:00 | 2019-023T00:00:00 | 2019-026T15:00:00 |
| 4 | 2019-045T09:00:00 | 2019-048T22:00:00 | 2019-041T08:00:00 | 2019-044T18:00:00 |
| 5 | 2019-094T04:00:00[a] | 2019-094T11:00:00[a] | 2019-087T06:00:00 | 2019-092T06:00:00 |
| 6 | 2019-144T12:00:00 | 2019-147T19:00:00 | 2019-139T02:00:00 | 2019-143T06:00:00 |
| 7 | 2019-171T05:00:00 | 2019-172T07:00:00 | 2019-162T10:00:00 | 2019-167T10:00:00 |
| 8 | 2019-197T15:00:00 | 2019-200T22:00:00 | 2019-186T18:00:00 | 2019-191T18:00:00 |

[a]Because this event is impulsive, lower energy particles arrive at the spacecraft later and so we average EPI-Lo fluxes from 2019-094T07:00:00 to 2019-094T13:00:00.